\documentclass[11pt]{article}
\usepackage{moriond}

\def\Journal#1#2#3#4{{#1} {\bf #2}, #3 (#4)}

\def\be{\begin{equation}}
\def\ee{\end{equation}}
\def\bea{\begin{eqnarray}}
\def\eea{\end{eqnarray}}

\newcommand{\Photo}{}

\begin{document}
\vspace*{4cm}
\title{THE ASTRI PROJECT: A MINI ARRAY OF DUAL-MIRROR SMALL CHERENKOV TELESCOPES FOR CTA}

\author{N.~La~Palombara$^{1}$,
G.~Agnetta$^{2}$,
L.A.~Antonelli$^{3}$,
D.~Bastieri$^{4}$,
G.~Bellassai$^{5}$,
M.~Belluso$^{5}$,
C.~Bigongiari$^{6a}$,
S.~Billotta$^{5}$,
B.~Biondo$^{2}$,
G.~Bonanno$^{5}$,
G.~Bonnoli$^{7}$,
P.~Bruno$^{5}$,
A.~Bulgarelli$^{8}$,
R.~Canestrari$^{7}$,
M.~Capalbi$^{2}$,
P.~Caraveo$^{1}$,
A.~Carosi$^{3}$,
E.~Cascone$^{9}$,
O.~Catalano$^{2}$,
M.~Cereda$^{7}$,
P.~Conconi$^{7}$,
V.~Conforti$^{8}$,
G.~Cusumano$^{2}$,
V.~De~Caprio$^{9}$,
A.~De~Luca$^{1}$,
A.~Di~Paola$^{3}$,
F.~Di~Pierro$^{6a}$,
D.~Fantinel$^{10}$,
M.~Fiorini$^{1}$,
D.~Fugazza$^{7}$,
D.~Gardiol$^{6b}$,
M.~Ghigo$^{7}$,
F.~Gianotti$^{8}$,
S.~Giarrusso$^{2}$,
E.~Giro$^{10}$,
A.~Grillo$^{5}$,
D.~Impiombato$^{2}$,
S.~Incorvaia$^{1}$,
A.~La~Barbera$^{2}$,
V.~La~Parola$^{2}$,
G.~La Rosa$^{2}$,
L.~Lessio$^{10}$,
G.~Leto$^{5}$,
S.~Lombardi$^{3}$,
F.~Lucarelli$^{3}$,
M.C.~Maccarone$^{2}$,
G.~Malaguti$^{8}$,
G.~Malaspina$^{7}$,
V.~Mangano$^{2}$,
D.~Marano$^{5}$,
E.~Martinetti$^{5}$,
R.~Millul$^{7}$,
T.~Mineo$^{2}$,
A.~Mist\'{o}$^{7}$,
C.~Morello$^{6a}$,
G.~Morlino$^{11}$,
M.R.~Panzera$^{7}$,
G.~Pareschi$^{7}$,
G.~Rodeghiero$^{10}$,
P.~Romano$^{2}$,
F.~Russo$^{2}$,
B.~Sacco$^{2}$,
N.~Sartore$^{1}$,
J.~Schwarz$^{7}$,
A.~Segreto$^{2}$,
G.~Sironi$^{7}$,
G.~Sottile$^{2}$,
A.~Stamerra$^{6a}$,
E.~Strazzeri$^{2}$,
L.~Stringhetti$^{1}$,
G.~Tagliaferri$^{7}$,
V.~Testa$^{3}$,
M.C.~Timpanaro$^{5}$,
G.~Toso$^{1}$,
G.~Tosti$^{12}$,
M.~Trifoglio$^{8}$,
P.~Vallania$^{6a}$,
S.~Vercellone$^{2}$,
V.~Zitelli$^{13}$ \\
\vspace*{0.5cm}}

\address{\scriptsize{$^{1}$ INAF - IASF Milano, Via E. Bassini 15, I-20133 Milano, Italy\\
$^{2}$ INAF - IASF Palermo, Via U. La Malfa 153, I-90146 Palermo, Italy\\
$^{3}$ INAF - Osservatorio Astronomico di Roma, Via Frascati 33, I-00040 Monte Porziocatone (RM), Italy\\
$^{4}$ Universit\'{a} di Padova, Dip. Fisica e Astronomia, Via Marzolo 8, I-35131 Padova, Italy\\
$^{5}$ INAF - Osservatorio Astrofisico di Catania, Via S. Sofia 78, I-95123 Catania, Italy\\
$^{6}$ INAF - Osservatorio Astrofisico di Torino: (a) Sede di Torino - Via P. Giuria 1, I-10125 Torino, Italy;\\
(b) Sede di Pino Torinese - Strada Osservatorio 20, I-10025 Pino Torinese (TO), Italy\\
$^{7}$ INAF - Osservatorio Astronomico di Brera, Via E. Bianchi 46, I-23807 Merate (LC), Italy\\
$^{8}$ INAF - IASF Bologna, Via P. Gobetti 101, I-40129 Bologna, Italy\\
$^{9}$ INAF – Osservatorio Astronomico di Capodimonte, Salita Moiariello 16, I-80131 Napoli, Italy\\
$^{10}$ INAF - Osservatorio Astronomico di Padova, Vicolo Osservatorio 5, I-35122 Padova, Italy\\
$^{11}$ INAF - Osservatorio Astrofisico di Arcetri, Largo E. Fermi 5, I-50125 Firenze, Italy\\
$^{12}$ Universit\'{a} di Perugia, Dip. Fisica, Via A. Pascoli, I-06123 Perugia, Italy\\
$^{13}$ INAF - Osservatorio Astronomico di Bologna, Via Ranzani 1, I-40127 Bologna, Italy}}

\maketitle\abstracts{ASTRI is a flagship project of the Italian Ministry of Education, University and Research, which aims to develop an end-to-end prototype of the CTA small-size telescope. The proposed design is characterized by a dual-mirror Schwarzschild-Couder configuration and a camera based on Silicon photo-multipliers, two challenging but innovative technological solutions which will be adopted for the first time on a Cherenkov telescope. Here we describe the current status of the project, the expected performance and the possibility to realize a mini-array composed by a few small-size telescopes, which shall be placed at the final CTA Southern Site.}

\section{The ASTRI project}\label{sec:project}

ASTRI (`Astrofisica con Specchi a Tecnologia Replicante Italiana') is a flagship project of the Italian Ministry of Education, University and Research, which, under the leadership of the Italian National Institute of Astrophysics (INAF), aims to develop the `replica' technology for mirrors and sensors for very-high energy (VHE) astrophysics. It is strictly linked to the Cherenkov Telescope Array (CTA, Actis et al. 2011~\cite{a}), since it is currently developing an end-to-end prototype of the Small-Size scale Telescope (SST) with wide field of view, aimed to observe the highest energy range (E $\sim$ 1-100 TeV) investigated by CTA.

The ASTRI prototype is characterized by two special features (Fiorini et al. 2012~\cite{b}): the optical system is designed in a dual-mirror configuration (SST-2M); the camera is formed by an array of Silicon photo-multipliers. The ASTRI SST-2M prototype is currently under construction and it will be tested on field: it is scheduled to start acquisition in 2014.

Beside the prototype, the ASTRI project aims to realize, in collaboration with CTA international partners, a mini-array of SST-2M telescopes. Among them, at least three ASTRI SST-2M telescopes are foreseen; they shall constitute, starting operation in 2016, the first seed of the CTA Observatory at its Southern site.
\vspace{-0.25cm}

\begin{figure}[t]
  \begin{center}
    \includegraphics[width=16cm]{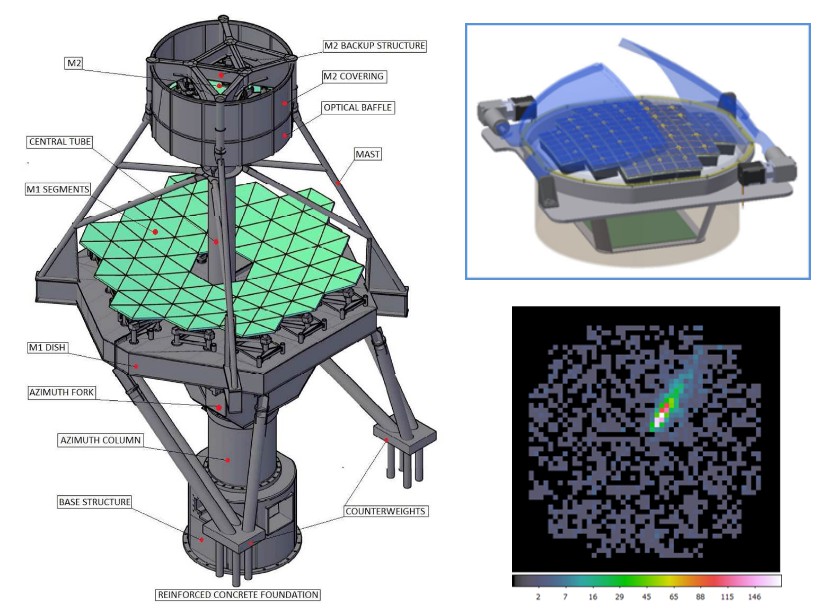}
%\begin{tabular}{c@{\hspace{1pc}}c}
%    \includegraphics[height=4.5cm]{structure.jpg} &
%    \includegraphics[height=4.5cm]{camera.jpg} \\
%    \includegraphics[height=4.5cm]{event.jpg}
%\end{tabular}
  \end{center}
\vspace{-0.5cm}
\caption{Layout of the ASTRI prototype (left panel) and of its camera (upper right panel) with lid opening. The lower right panel shows how a Cherenkov event would be seen by the ASTRI camera (see text for details).}
\label{fig1}
\vspace{-0.25cm}
\end{figure}

\section{The ASTRI prototype}\label{sec:prototype}

The ASTRI SST-2M, for the first time for a Cherenkov telescope, will adopt a dual-mirror Schwarzschild-Couder (SC) optical design (Vassiliev et al. 2007~\cite{c}), which is characterized by a wide field of view (FoV = 9.6$^\circ$ in diameter) and a compact optical configuration (f-number f/0.5). In this way it will be possible to use a light and compact camera based on Silicon photo-multipliers, with a logical pixel size of 6.2 mm $\times$ 6.2 mm, corresponding to an angular size of 0.17$^\circ$. Figure~\ref{fig1} (left panel) shows the telescope layout, whose mount exploits the classical altazimuthal configuration, and which is fully compliant with the CTA requirements for the SST array. The ASTRI SST-2M prototype will be placed at Serra La Nave, on the Etna Mountain near Catania, at the INAF `M.G. Fracastoro' observing station 1735 m a.s.l. (Maccarone 2011~\cite{d}); it will begin data acquisition in 2014.

%\subsection{The optical design}\label{subsec:design}

\smallskip
\noindent
\underline{{\bf The Optical Design.}} The proposed layout (Canestrari et al. 2011~\cite{e}) is characterized by a wide-field aplanatic optical configuration: it is composed by a segmented primary mirror made of three different types of segments, a concave secondary mirror, and a convex focal surface. The design has been optimized in order to ensure, over the entire FoV, a light concentration higher than 80 \% within the angular size of the pixels. The telescope design is compact, since the primary mirror (M1) and the secondary mirror (M2) have a diameter of 4.3 m and 1.8 m, respectively, and the primary-to-secondary distance is 3 m. The SC optical design has an f-number f/0.5, a plate scale of 37.5 mm/$^\circ$, a logical pixel size of approximately 0.17$^\circ$, an equivalent focal length of 2150 mm and a FoV of 9.6$^\circ$ in diameter. The mean value of the active area is $\sim$ 6.5 m$^2$, which takes into account the segmentation of M1, the obscuration of M2, the obscuration of the camera, the reflectivity of the optical surfaces (as a function of the wavelength and incident angle), the losses due to the protection window of the camera and the efficiency of the silicon detectors as function of the incident angles (ranging from 25$^\circ$ to 72$^\circ$).

%\subsection{The Mirrors}\label{subsec:mirrors}

\smallskip
\noindent
\underline{{\bf The Mirrors.}} The primary mirror is composed by 18 hexagonal segments, with an aperture of 849 mm face-to-face; the central segment is not used because it is completely obstructed by the secondary mirror. According to their distance from the optical axis, there are three different types of segments, each having a specific surface profile. In order to perform the correction of the tilt misplacements, each segment will be equipped with a triangular frame with two actuators and one fixed point. The secondary mirror is monolithic and has a curvature radius of 2200 mm and a diameter of 1800 mm. It will be equipped with three actuators, where the third actuator will provide the piston/focus adjustment for the entire optical system. For both the segments of the primary mirror and the secondary mirror the reflecting surface is obtained with a Vapor Deposition of a multilayer of pure dielectric material, a technology approach developed at the INAF Brera observatory.

%\subsection{The Camera}\label{subsec:camera}

\smallskip
\noindent
\underline{{\bf The Camera.}} The SC optical configuration allows us to design a compact and light camera. Currently, the ASTRI camera has a dimension of about 500 mm $\times$ 500 mm $\times$ 500 mm, including the mechanics and the interface with the telescope structure, for a total weight of $\sim$ 50 kg. Such small detection surface, in turn, requires a spatial segmentation of a few square millimeters to be compliant with the imaging resolving angular size. In addition, the light sensor shall offer a high photon detection sensitivity in the wavelength range between 300 and 700 nm and a fast temporal response. In order to be compliant with these requirements, we selected the Hamamatsu Silicon Photomultiplier (SiPM) S11828-3344M (Hamamatsu 2011~\cite{f}). The `unit' provided by the manufacturer is the physical aggregation of 4 $\times$ 4 pixels (3 mm $\times$ 3 mm each pixel), while the logical aggregation of 2 $\times$ 2 pixels is a `logical pixel'; its size of 6.2 mm $\times$ 6.2 mm corresponds to 0.17$^\circ$. In order to cover the full FoV, we adopt a modular approach: we aggregate 4 $\times$ 4 units in a Photon Detection Module (PDM) and, then, use 37 PDMs to cover the full FOV. The advantage of this design is that each PDM is physically independent of the others, allowing maintenance of small portions of the camera. To fit the curvature of the focal surface, each PDM is appropriately tilted with respect to the optical axis.

The camera is equipped with a light-tight two-petal lid (Figure~\ref{fig1}, upper right) in order to prevent accidental sunlight exposure of its SiPM detectors. Eventually, Figure~\ref{fig1} (lower right panel) shows how the ASTRI camera would `see' a Cherenkov event; the example refers to an on-axis simulated event for a primary gamma-ray with 10 TeV energy, a core distance of about 143 m, and embedded in a night-sky background of 1.9 $\times$ 10$^{-12}$ photons m$^{-2}$ s$^{-1}$ sr$^{-1}$; the color table indicates the number of photoelectrons registered in each logical pixel.

%\subsection{The Prototype Expected Performance}\label{sec:performance}

\begin{figure}[t]
\begin{center}
\includegraphics[height=6cm]{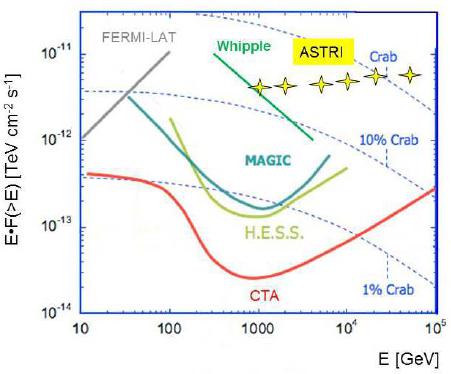}
\includegraphics[height=6cm]{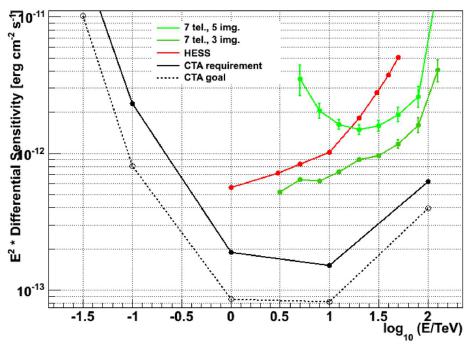}
\end{center}
\vspace{-0.5cm}
\caption{Expected sensitivity as a function of the energy for the ASTRI SST-2M prototype (\textit{left panel}, yellow stars) and mini-array (\textit{right panel}, green lines), computed at 5 $\sigma$ and 50 hr of observation.}
\label{fig2}
\end{figure}

\smallskip
\noindent
\underline{{\bf The Prototype Expected Performance.}} Although the ASTRI prototype will mainly be a technological prototype, it should be able to perform also scientific observations. Based on the foreseen maximum sensitivity, a source flux of 1 Crab at E $>$ 2 TeV should be detectable at 5 $\sigma$ confidence level in some hours, while a few tens of hours should be necessary to obtain a comparable detection at E $>$ 10 TeV (Vallania et al. 2012~\cite{g}). In this way we would obtain the first Crab observations with a Cherenkov telescope adopting a Schwarschild-Couder optical design and a SiPM camera. Figure~\ref{fig2} (left panel) shows the expected ASTRI prototype sensitivity (yellow stars, computed at 5 $\sigma$ confidence level and 50 hr of observation) compared to those of Fermi-LAT (for one-year integration) and of a few Image Atmospheric Cherenkov Telescopes.

\section{The ASTRI SST-2M Mini-Array}

The ASTRI Project aims to realize also a mini-array of a few SST-2M telescopes, which shall be placed at the CTA Southern Site and start operations in 2016. Preliminary Monte Carlo simulations (Di Pierro et al. 2012~\cite{h}) yield an improvement in sensitivity that, for 7 telescopes at an optimized distance of 250-300 m, could be a factor 1.5 at 10 TeV w.r.t. H.E.S.S. (Figure~\ref{fig2}, right panel). The ASTRI SST-2M mini-array will be able to study in great detail sources with a flux of a few 10$^{-12}$ erg cm$^{-2}$ s$^{-1}$ at 10 TeV, with an angular resolution of a few arcmin and an energy resolution of about 10-15 \%. Moreover, thanks to the array approach, it will be possible to verify the wide FoV performance to detect very high energy showers with the core located at a distance up to 500 m, to compare the mini-array performance with the Monte Carlo expectations (by means of deep observations of few selected targets), and to perform the first CTA science, with its first solid detections during the first year of operation.

The Mini-array will observe prominent sources such as extreme blazars (1ES 0229+200), nearby well-known BL Lac objects (MKN 421 and MKN 501) and radio-galaxies, galactic pulsar wind nebulae (Crab Nebula, Vela-X), supernovae remnants (Vela-junior, RX J1713.7-3946) and microquasars (LS 5039), as well as the Galactic Center. In this way it will be possible to investigate the electron acceleration and cooling, to study the relativistic and non relativistic shocks, to search for cosmic-ray (CR) Pevatrons, to study the CR propagation and the impact of the extragalactic background light on the spectra of the nearby sources.

\section*{Acknowledgments}
\vspace{-0.25cm}

This work was partially supported by the ASTRI Flagship Project financed by the Italian Ministry of Education, University, and Research (MIUR) and lead by the Italian National Institute of Astrophysics (INAF).

\end{document}